\documentclass[preprint,aps,amsmath,byrevtex,prd,titlepage,
nofootinbib]{revtex4-2}

\usepackage{dcolumn}
\usepackage{amsmath}
\usepackage{amssymb}
\usepackage{bm}
\usepackage{hyperref}
\usepackage{graphicx}
\usepackage{slashed}

\newcommand{\beq}{\begin{equation}}
\newcommand{\eeq}{\end{equation}}
\newcommand{\eq}[1]{Eq.~(\ref{#1})}

\begin{document}

\title{Three-Loop Corrections to Lamb Shift in Muonium and Positronium}

\author {Michael I. Eides}
\altaffiliation[Also at ]{the Petersburg Nuclear Physics Institute,
Gatchina, St.Petersburg 188300, Russia}
\email[Email address: ]{meides@g.uky.edu, eides@thd.pnpi.spb.ru}
\affiliation{Department of Physics and Astronomy,
University of Kentucky, Lexington, KY 40506, USA}
\author{Valery A. Shelyuto}
\email[Email address: ]{shelyuto@vniim.ru}
\affiliation{D. I.  Mendeleyev Institute for Metrology,
St.Petersburg 190005, Russia}

\begin{abstract}
We calculate hard spin-independent three-loop radiative corrections to energy levels in muonium and positronium which are due to radiative insertions in two-photon exchange diagrams. These corrections could be relevant for the new generation of precise $1S-2S$ and $2S-2P$  measurements in muonium and positronium.
\end{abstract}

\maketitle

\section{Introduction}
Muonium ($Mu=\mu^+e^-$) and positronium ($Ps=e^+e^-$) are purely electrodynamic bound states which admit precise measurements and calculations of transition frequencies. For many years main emphasis was on the ground state hyperfine splitting, see experimental results for muonium in \cite{mbb1982,lbd1999,Kanda:2020mmc,MuSEUM:2020mzm}, for positronium in \cite{apmghb1975,apmj1983,Ritter:1984mqy,Sasaki:2010mg,Ishida:2013waa,Heiss:2018jbl} and references therein. Calculations of high order corrections to hyperfine splitting remain an active field of research. One can find some recent theoretical results of order $\alpha^2(Z\alpha)^5(m/M)m$ for muonium in \cite{Eides:2013yxa,Eides:2014nra,Eides:2014xea}, see also reviews \cite{Eides:2000xc,Eides:2007exa,Tiesinga:2021myr}. Hyperfine splitting in muonium serves as the best source for extracting a precise  value of the electron-muon mass ratio \cite{Eides:2018rph}. New contributions to hyperfine splitting in positronium of order $\alpha^7m$  were calculated recently in \cite{Baker:2014sua,Adkins:2014dva,Eides:2014nga,Adkins:2014xoa,Eides:2015nla,Adkins:2015jia,Adkins:2015dna,Adkins:2016btu,Eides:2017uoy}, see also reviews \cite{Karshenboim:2003vs,Karshenboim:2005iy,Adkins:2018lvj}.

Transition frequencies $1S-2S$ \cite{Chu:1988,Maas:1994,Meyer:1999cx,Fan:2013qpk} and the classical Lamb shift $2S-2P$ \cite{Oram:1983sd,Woodle:1990ky} in muonium which were measured some time ago were somewhat on back burner for a while. In recent years a lot of experimental efforts shifted to these energy intervals. New generation of $1S-2S$ measurements in muonium is currently planned \cite{Crivelli:2018vfe,Ohayon:2021dec,yksu2018}. The goal of the Phase 1 of the MU-MASS experiment at PSI is to reduce the experimental uncertainty to below   $<100$ kHz (40 ppt), and at Phase 2 it is planned to reduce it below  10 kHz (4 ppt) \cite{Crivelli:2018vfe}. The goal of the J-PARC experiment is to achieve experimental uncertainty about 10 kHz (4 ppt) \cite{yksu2018}. This is a one thousand times improvement in comparison with the previous measurements \cite{Crivelli:2018vfe}. The classical Lamb shift in muonium $2S-2P$  was recently measured \cite{Ohayon:2021qof} to be $1047.2(2.3)_{stat}(1.1)_{syst}$ MHz which is an order of magnitude smaller uncertainty than the best previous measurement \cite{Woodle:1990ky}. The goal of this ongoing experiment is to reduce the experimental uncertainty to about a few tens of kHz.

Transition frequencies $1S-2S$ in positronium were measured long time ago \cite{Chu:1984zz,Fee:1993zz}, uncertainty achieved in the last experiment is 2.4 ppb. New experiments are ongoing at ETH Zurich and at UC Riverside \cite{Crivelli:2016fjw,Mills:2016} and at the University College London \cite{Cassidy:2018tgq}. The goal is to reduce the experimental uncertainty to about 0.5 ppb \cite{Crivelli:2016fjw}. Results of a precise measurement of the fine structure in positronium  were recently reported \cite{Gurung:2020hms}.

Muonium atom is similar to hydrogen and purely quantum electrodynamic corrections in both cases are the same. Nonrecoil radiative corrections to hydrogen energy levels can be used for muonium as well. The difference between hydrogen and muonium arises in consideration of recoil and radiative-recoil corrections. These corrections in hydrogen strongly depend on the proton structure, do not reduce to pure QED, and require account for strong interactions. This makes the hydrogen problem more challenging and reduces theoretical accuracy of these corrections. For example, theory of hyperfine splitting in the ground state of hydrogen has the relative theoretical uncertainty about 1 ppm \cite{Tomalak:2018uhr}, while the theoretical uncertainty of the similar HFS in muonium is about 15 ppb \cite{Eides:2018rph} and admits further reduction. High accuracy in muonium is achieved because due to absence of the strong interaction effects higher order spin-dependent radiative-recoil contributions admit purely electrodynamic calculations. Positronium is also a purely electromagnetic bound state, which admits high precision calculations of energy levels.

\begin{figure}[h!]
\begin{center}
\includegraphics[height=2.5 cm]{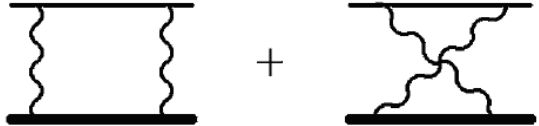}
\end{center}
\caption{Skeleton diagrams.}
\label{skeleton}
\end{figure}

Inspired by the  experimental progress on measurements of  $1S-2S$ and $2S-2P$ transitions in muonium and positronium we calculate hard three-loop spin-independent contributions to the energy levels. All corrections  considered below arise by insertions of radiative corrections in the skeleton diagrams in Fig.~\ref{skeleton}. These corrections are similar to the respective spin-dependent corrections to hyperfine splitting and are generated by the same sets of gauge invariant diagrams \cite{Eides:2009dp,Adkins:2014dva,Eides:2014nga,Adkins:2014dva,Eides:2014nga}. In the case of muonium three-loop nonrecoil spin-independent contributions generated by these diagrams were calculated long time ago \cite{Eides:1992gr}, and we calculate below respective  radiative-recoil corrections of order $\alpha^2(Z\alpha)^5 (m/M)m$. The skeleton integral for the recoil corrections obtained by subtraction of the nonrecoil contribution has the form \cite{Eides:1994mm}

\beq                       \label{skelrec}
\begin{split}
\Delta E^{(Mu)}_{skel-rec}
&=\frac{16(Z\alpha)^5m}{\pi n^3(1-\mu^2)}\left(\frac{m_r}{m}\right)^3\int_0^\infty
\frac{kdk}{(k^2+\lambda^2)^2}
\Biggl[\mu\sqrt{1+\frac{k^2}{4}}\left(\frac{1}{k}+\frac{k^3}{8}\right)\\
&-\sqrt{1+\frac{\mu^2k^2}{4}}\left(\frac{1}{k}+\frac{\mu^4k^3}{8}\right)
-\frac{\mu k^2}{8}\left(1+\frac{k^2}{2}\right)
+\frac{\mu^3k^2}{8}\left(1+\frac{\mu^2k^2}{2}\right )
+\frac{1}{k}\Biggr]\delta_{l0},
\end{split}
\eeq

\noindent
where $m$ and $M$ are the electron and muon masses, respectively, $m_r=mM/(m+M)$ is the reduced mass, $\mu=m/M$, $\lambda$ is an auxiliary mass of the exchanged photon to be omitted below, $n$ and $l$ are the principal quantum number and the orbital momentum, respectively, and the dimensionless integration
momentum is measured in units of the electron mass.

Corrections to the Lamb shift of order $\alpha^7m$  in positronium considered below are obtained by the radiation insertions in the same two-photon exchange diagrams in Fig.~\ref{skeleton}. The mass ratio of the constituents in positronium is one and separation into recoil and nonrecoil corrections does not make much sense. The skeleton integral for the two-photon exchange diagrams in positronium has the form \cite{Eides:1994mm}

\beq                       \label{skelrecpos}
\Delta E^{(Ps)}_{skel-rec}
=\frac{2\alpha^5m}{\pi n^3}\int_0^\infty dk\left[\frac{k^2}{8 \sqrt{k^2+4}}+\frac{3}{8 \sqrt{k^2+4}}-\frac{1}{\sqrt{k^2+4}
   k^4}-\frac{k}{8}-\frac{1}{8 k}\right]\delta_{l0}.
\eeq

\section{Calculations of Gauge Invariant Contributions}

\subsection{Diagrams with One-Loop Polarization Insertions}

\subsubsection{Muonium}

\begin{figure}[h!]
\begin{center}
\includegraphics[height=3 cm]{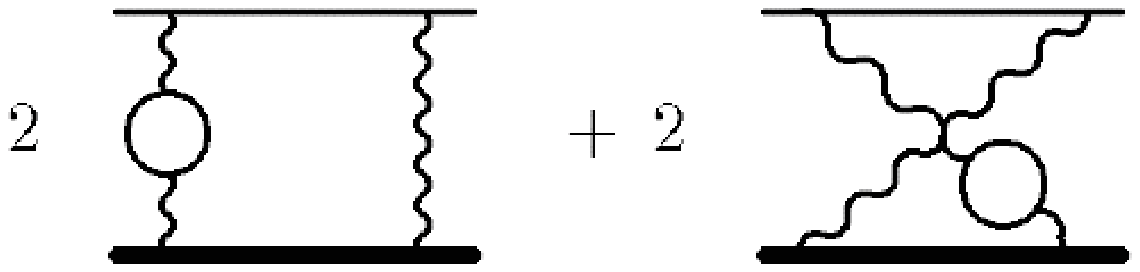}
\end{center}
\caption{Graphs with one one-loop polarization insertions.}
\label{onepol}
\end{figure}

Radiative-recoil corrections generated by the diagrams in Fig.~\ref{onepol} can be obtained from the skeleton expression in \eq{skelrec} by the substitution

\beq  \label{subprevor}
\frac{1}{k^2}\rightarrow 2\left(\frac{\alpha}{\pi}\right)I_{1}(k),
\eeq

\noindent
where

\beq
{I_1(k)}= \int_0^1 dv \frac{v^2(1-v^2/3)}{4+(1-v^2)k^2}.
\eeq

\noindent
is the one-loop polarization operator.

This contribution was calculated long time ago \cite{Eides:1994mm}

\beq
\Delta E=\left[\left(\frac{2\pi^2}{9}-\frac{70}{27}\right)-\frac{3\pi^2}{16}\frac{m}{M}\right]\frac{\alpha(Z\alpha)^5}{\pi^2n^3}\frac{m}{M}
\biggl(\frac{m_r}{m}\biggr)^3m \delta_{l0},
\eeq

\noindent
where we restored  correction of the relative order $(m/M)^2$ omitted in \cite{Eides:1994mm}.

The spin-independent radiative-recoil contribution of the next order in $\alpha/\pi$

\beq                       \label{radrec1}
\begin{split}
\Delta E^{(Mu)}_1
&=\frac{48(Z\alpha)^5m}{\pi n^3(1-\mu^2)}\left(\frac{\alpha}{\pi}\right)^2\left(\frac{m_r}{m}\right)^3\int_0^\infty
kdk I_1^2(k)
\Biggl\{\mu\sqrt{1+\frac{k^2}{4}}\left(\frac{1}{k}+\frac{k^3}{8}\right)\\
&-\sqrt{1+\frac{\mu^2k^2}{4}}\left(\frac{1}{k}+\frac{\mu^4k^3}{8}\right)
-\frac{\mu k^2}{8}\left(1+\frac{k^2}{2}\right)
+\frac{\mu^3k^2}{8}\left(1+\frac{\mu^2k^2}{2}\right )
+\frac{1}{k}\Biggr\}\delta_{l0},
\end{split}
\eeq

\noindent
generated by the diagrams in Fig.~\ref{twoonepol}\footnote{Diagrams with the crossed exchange photons are omitted in this figure and other figures below.}  can be obtained from the skeleton expression in \eq{skelrec} by the substitution

\beq         \label{onelsub}
\frac{1}{k^2}\rightarrow 3\left(\frac{\alpha}{\pi}\right)^2k^2I^2_{1}(k),
\eeq

\noindent
where 3 is a combinatorial factor which arises due to three ways to insert polarization operator in the skeleton graphs.

\begin{figure}[h!]
\begin{center}
\includegraphics[height=2 cm]{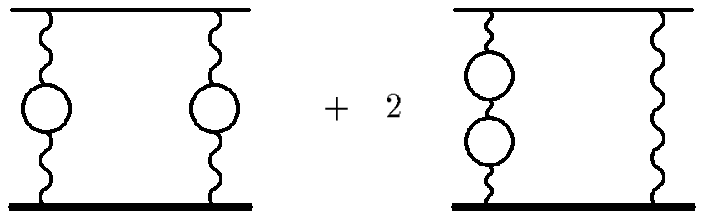}
\end{center}
\caption{Graphs with two one-loop polarization insertions. }
\label{twoonepol}
\end{figure}

\noindent
The integral in \eq{radrec1} contains corrections of all orders in the mass ratio and can be easily calculated numerically with an arbitrary accuracy

\beq \label{numradrec1}
\Delta E^{(Mu)}_1=0.959540854(3)\ldots\frac{\alpha^2(Z\alpha)^5}{\pi^3n^3}\frac{m}{M}\delta_{l0}.
\eeq

\noindent
One can expand the integral in \eq{radrec1} up to the first order in $\mu=m/M$ and obtain an analytic result
\beq \label{deltae1}
\begin{split}
\Delta E^{(Mu)}_1&\approx
\frac{48(Z\alpha)^5m}{\pi n^3}\left(\frac{\alpha}{\pi}\right)^2\left(\frac{m_r}{m}\right)^3\mu\int_0^\infty
kdk I_1^2(k)\Biggl[\sqrt{1+\frac{k^2}{4}}\biggl(\frac{1}{k}+\frac{k^3}{8}\biggr)
- \frac{k^2}{8}\biggl(1+\frac{k^2}{2}\biggr)\Biggr]\delta_{l0}\\
&=\left(\frac{1541}{486}-\frac{172 }{2835}\pi^2 - \frac{4}{3}\zeta{(3)}\right)\frac{\alpha^2(Z\alpha)^5}{\pi^3n^3}\frac{m}{M}
\biggl(\frac{m_r}{m}\biggr)^3m\delta_{l0}\\
&=0.9692\ldots\frac{\alpha^2(Z\alpha)^5}{\pi^3n^3}\frac{m}{M}
\biggl(\frac{m_r}{m}\biggr)^3m \delta_{l0}.
\end{split}
\eeq

\noindent
We can calculate higher order terms in the expansion of the integral in \eq{radrec1} in $\mu$ which restore  agreement between the numerical factors in \eq{numradrec1} and \eq{deltae1}.

\subsubsection{Positronium}

Corrections of order $\alpha^7 m$ generated by the diagrams in Fig.~\ref{onepol} are obtained from the skeleton expression in \eq{skelrecpos} by the substitution in \eq{onelsub}. After calculations we obtain

\beq \label{positrvac1}
\Delta E^{(Ps)}_1=\left(-\frac{\zeta (3)}{6}+\frac{1709}{3888}-\frac{11 \pi ^2}{405}\right)\frac{\alpha^7m}{\pi^3n^3}m\delta_{l0}
=-0.028848\ldots\frac{\alpha^7m}{\pi^3n^3}\delta_{l0}.
\eeq

\subsection{Diagrams with Two-Loop Polarization Insertions}

\subsubsection{Muonium}

\begin{figure}[h!]
\begin{center}
\includegraphics[height=2 cm]{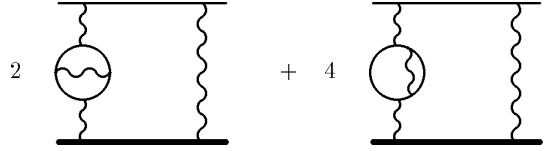}
\end{center}
\caption{Graphs with two-loop polarization insertions. }
\label{twoloppol}
\end{figure}

The spin-independent radiative-recoil contribution  generated by the diagrams in Fig.~\ref{twoloppol} can be obtained from the skeleton expression in \eq{skelrec} by the substitution

\beq         \label{sub2}
\frac{1}{k^2}\rightarrow 2\left(\frac{\alpha}{\pi}\right)^2I_{2}(k),
\eeq

\noindent
where the irreducible two-loop polarization has the form \cite{Kallen:1955,Schwinger:1973}

\beq
I_2(k) =\int_0^{1}{dv}\frac{\frac34\cdot v^2\Bigl(1-\frac{v^2}{3}\Bigr)+ R(v)}{4+(1-v^2)k^2},
\eeq

\noindent
and

\beq \label{rvpol}
\begin{split}
R(v)&= \frac{2}{3} v \biggl\{(3-v^2)(1+v^2)\biggl[\mbox{Li}_2 \biggl(
-\frac{1-v}{1+v}\biggr)
+ 2\mbox{Li}_2 \biggl(\frac{1-v}{1+v}\biggr)
+\frac {3}{2} \ln{\frac{1+v}{1-v}} \ln{\frac{1+v}{2}}
\\
&- \ln{\frac{1+v}{1-v}} \ln{v} \biggr]
+ \biggl[\frac{11}{16}(3-v^2)(1+v^2) + \frac{v^4}{4}\biggr]
\ln{\frac{1+v}{1-v}}
+\biggl[\frac{3}{2}v(3-v^2)\ln{\frac{1-v^2}{4}}\\
&
- 2v(3-v^2)\ln{v} \biggr]+ \frac{3}{4} v(1-v^2)\biggl\}.
\end{split}
\eeq

Like the skeleton in \eq{skelrec} at $\lambda=0$ the momentum integral with the two-loop polarization insertion is linearly infrared divergent. In a more accurate calculation the divergence would be cutoff at the characteristic wave function momentum $k\sim Z\alpha$. This divergence reflects existence of a contribution of a lower order in $Z\alpha$, for more details see \cite{Eides:2000xc,Eides:2007exa,Eides:1994mm}.  This contribution is well known and to get rid of its remnants we simply subtract from the integrand the infrared divergent term

\beq
\frac{\mu}{k^2}I_2(0)=\frac{41}{162}\frac{\mu}{k^2}.
\eeq

After subtraction the fully convergent expression for the contribution of the two-loop polarization in  Fig.~\ref{twoloppol} has the form

\beq                       \label{toloopallor}
\begin{split}
\Delta E^{(Mu)}_2
&=\frac{32(Z\alpha)^5m}{\pi n^3(1-\mu^2)}\left(\frac{\alpha}{\pi}\right)^2\left(\frac{m_r}{m}\right)^3\int_0^\infty
\frac{dk}{k}
\Biggl\{I_2(k)\Biggl[\mu\sqrt{1+\frac{k^2}{4}}\left(\frac{1}{k}+\frac{k^3}{8}\right)\\
&-\sqrt{1+\frac{\mu^2k^2}{4}}\left(\frac{1}{k}+\frac{\mu^4k^3}{8}\right)
-\frac{\mu k^2}{8}\left(1+\frac{k^2}{2}\right)
+\frac{\mu^3k^2}{8}\left(1+\frac{\mu^2k^2}{2}\right )
+\frac{1}{k}\Biggr]-\frac{41}{162}\frac{\mu}{k}\Biggr\}\delta_{l0}.
\end{split}
\eeq

\noindent
Like in the case of one-loop polarization above the integral in \eq{toloopallor} contains corrections of all orders in the mass ratio and can be easily calculated numerically with an arbitrary accuracy

\beq \label{numradrec2}
\Delta E^{(Mu)}_2=-3.133412(3)\ldots \frac{\alpha^2(Z\alpha)^5}{\pi^3 n^3}\frac{m}{M}\left(\frac{m_r}{m}\right)^3m\delta_{l0}.
\eeq

\noindent
One can expand the integral in \eq{toloopallor} up to the first order in $\mu=m/M$ and obtain an analytic result

\beq \label{deltae2}
\begin{split}
\Delta E^{(Mu)}_2&\approx
\frac{32(Z\alpha)^5m}{\pi n^3}\left(\frac{\alpha}{\pi}\right)^2\left(\frac{m_r}{m}\right)^3\mu\int_0^\infty{\frac{dk}{k}}
\Biggl\{I_2(k)\Biggl[\sqrt{1+\frac{k^2}{4}}\biggl(\frac{1}{k}+\frac{k^3}{8}\biggr)\\
&-\frac{k^2}{8}~\biggl(1+\frac{k^2}{2}\biggr)\Biggr]
- \frac{41}{162k}\Biggr\}\delta_{l0}
=\Biggl(\frac{6589}{7560} + \frac{145756 \pi^2}{99225} +  \frac{7\pi^4}{270}
- \frac{296\pi^2}{315}\ln{2}\\
&+  \frac{4\pi^2}{9}\ln^2{2} -  \frac{4}{9}\ln^4{2}
- \frac{32}{3}\mbox{Li}_4\biggl(\frac{1}{2}\biggr)  -     \frac{11597}{1260}\,\zeta{(3)}\Biggr)
\frac{\alpha^2(Z\alpha)^5}{\pi^3 n^3}\frac{m}{M}\left(\frac{m_r}{m}\right)^3m\delta_{l0}\\
&=-3.1121\ldots\frac{\alpha^2(Z\alpha)^5}{\pi^3 n^3}\frac{m}{M}\left(\frac{m_r}{m}\right)^3m\delta_{l0}.
\end{split}
\eeq

\noindent
As in the case of one-loop polarizations we can calculate higher order terms in the expansion of the integral in \eq{toloopallor} in $\mu$ which restore  agreement between the numerical factors in \eq{numradrec2} and \eq{deltae2}.

\subsubsection{Positronium}

The  Lamb shift contribution of order $\alpha^7m$ in positronium generated by the diagrams in Fig.~\ref{twoloppol}  is obtained from the skeleton expression in \eq{skelrecpos} by the substitution in \eq{sub2}. Like in the case of muonium the integral with the two-loop polarization insertion is linearly infrared divergent. This divergence arises because the integral after substitution contains also the contribution of order $\alpha^6m$ which should be subtracted. After subtraction the convergent integral has the form

\beq
\Delta E^{(Ps)}_2=\frac{4\alpha^7m}{\pi^3n^3}\int_0^\infty dk I_2(k)\left[\frac{k^4}{8 \sqrt{k^2+4}}+\frac{3k^2}{8 \sqrt{k^2+4}}-\frac{k^3+k}{8}-\frac{1}{\sqrt{k^2+4} k^2}+\frac{41}{324 k^2}\right]\delta_{l0}.
\eeq

\noindent
Calculating this integral we obtain

\beq \label{positrvac2}
\begin{split}
\Delta E^{(Ps)}_2&=\Biggl(-\frac{4 \text{Li}_4\left(\frac{1}{2}\right)}{3}-\frac{17921 \zeta
   (3)}{10080}+\frac{26347}{60480}+\frac{311233 \pi^2}{793800}\\
&+\frac{7 \pi ^4}{2160}+\frac{1}{18}\pi^2\ln^22-\frac{\ln ^42}{18}-\frac{76}{315} \pi^2\ln 2\Biggr)\frac{\alpha^7m}{\pi^3n^3}\delta_{l0}\\
&=0.393966\ldots\frac{\alpha^7m}{\pi^3n^3}\delta_{l0}.
\end{split}
\eeq

\subsection{Diagrams with One-Loop Electron Factor}

\begin{figure}[h!]
\begin{center}
\includegraphics[height=2 cm]{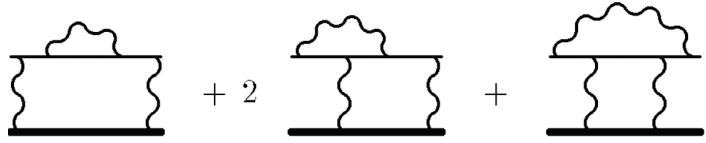}
\end{center}
\caption{Electron-line radiative-recoil corrections.}
\label{elfact}
\end{figure}

The contribution to the spin-independent energy shift generated by the diagrams in Fig.~\ref{elfact} is given by the integral \cite{Eides:2000kj}

\beq  \label{general}
\Delta E=-\frac{(Z\alpha)^5}{\pi n^3}{m_r}^3
\int {\frac{d^4 k}{i\pi^2 k^4}} \frac{1}{4} Tr \Bigl[(1 + \gamma_0 )L_{\mu \nu} \Bigr]\:
\frac{1}{4} Tr \Bigl[(1 + \gamma_0 )H_{\mu \nu} \Bigr]~\delta_{l0},
\eeq

\noindent
where $L_{\mu \nu}$ and $H_{\mu \nu}$ are the electron and muon factors, respectively.

\noindent
The electron factor is equal to the sum of the self-energy, vertex, and
spanning photon insertions in the electron line

\beq
L_{\mu \nu}= L_{\mu \nu}^{\Sigma} + 2L_{\mu \nu}^{\Lambda}
+ L_{\mu \nu}^{\Xi},
\eeq

\noindent
and the heavy-line muon factor is given by the expression

\beq
H_{\mu \nu}=\gamma_{\mu} \frac{\hat{P} + \hat{k} + M}
{k^2 + 2Mk_0 + i0} \gamma_{\nu}+
\gamma_{\nu}  \frac{\hat{P} - \hat{k} + M}{k^2 - 2Mk_0 +
i0}\gamma_{\mu},
\eeq

\noindent
where $P=(M,{\bf 0})$ is the momentum of the muon.

The expression for the energy shift in \eq{general} contains both recoil and nonrecoil contributions of order
$\alpha(Z\alpha)^5m$. The nonrecoil correction is well known from the  early days of quantum electrodynamics and subtracting it and preserving only linear in mass ratio contribution we obtain the integral for the respective radiative-recoil contribution \cite{Eides:2000kj}

\beq \label{ordermm}
\begin{split}
\Delta E =&\frac{(Z\alpha)^5}{\pi n^3}
\frac{m_r^3}{M}
\int {\frac{d^4 k}{i\pi^2 k^4}}
\frac{1}{4} Tr \{(1 + \gamma_0 )
L_{\mu \nu}\}\\
&
\times\left[k^2 g_{\mu 0} g_{\nu 0}
\wp \left(\frac{1}{k_0^2}\right)
- (g_{\mu 0} k_{\nu} + g_{\nu 0} k_{\mu})
\frac{1}{k_0} + g_{\mu \nu} \right]\delta_{l0}.
\end{split}
\eeq

\noindent
Here $\wp ({1}/{k_0^2}$) is a slightly nonstandard principal value integration prescription, for its precise definition and properties see \cite{Eides:2000kj}.

\begin{figure}[h!]
\begin{center}
\includegraphics[height=2 cm]{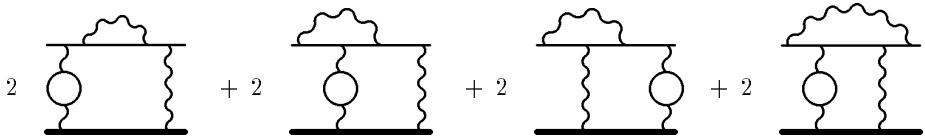}
\end{center}
\caption{Graphs with radiative insertions in the electron line and one-loop polarization in the exchanged photons. }
\label{ee}
\end{figure}

The expression for the spin-independent radiative-recoil contribution of order $\alpha^2(Z\alpha)^5(m/M)m$ generated by the diagrams in Fig.~\ref{ee} can be obtained from the Wick rotated integral in \eq{ordermm} by the substitution in \eq{subprevor}.

Let us mention that the expression for the energy shift in \eq{ordermm} is linearly infrared divergent like $1/\gamma$ where $\gamma$ is an auxiliary infrared cutoff in integration over $k$. This linear infrared divergence arises because the expressions for the energy shifts in \eq{general} and \eq{ordermm} contain not only corrections of  order $\alpha(Z\alpha)^5$ but also the corrections of the previous order in $Z\alpha$.  The divergent contribution was subtracted in \cite{Eides:2000kj} in order to obtain an integral representation for the contribution of order $\alpha(Z\alpha)^5$. The Wick rotated integral obtained from \eq{ordermm} after the substitution in \eq{subprevor} contains only linear in the mass ratio contributions of order $\alpha^2(Z\alpha)^5$ and no such subtraction is necessary.

Next we calculate the contributions to the energy shift of the four diagrams in Fig.~\ref{ee} in the Yennie gauge

\beq \label{VP-eLine-01}
\Delta E^{(Mu)}_3=(J_{\Sigma P} + 2J_{\Lambda P} + J_{\Xi P})\frac{\alpha^2(Z\alpha)^5}{\pi^3n^3}\frac{m}{M}
\left(\frac{m_r}{m}\right)^3m\delta_{l0}.
\eeq

\noindent
Calculations are similar to the ones in  \cite{Eides:2000kj}, and we obtain

\beq
J_{\Sigma P}=2.2619(1),\quad 2J_{\Lambda P}=-14.948(1),\quad J_{\Xi P}=3.4292(1).
\eeq

\noindent
Finally, total contribution to the Lamb shift of the diagrams in Fig.~\ref{elfact} is

\beq \label{elfVP}
\Delta E^{(Mu)}_3=-9.2569(2)\frac{\alpha^2(Z\alpha)^5}{\pi^3n^3}\frac{m}{M}
\left(\frac{m_r}{m}\right)^3m\delta_{l0}.
\eeq

\section{Summary of Results}

Collecting the results in \eq{numradrec1}, \eq{numradrec2} and \eq{elfVP} we obtain the total spin-independent radiative-recoil contribution of order $\alpha^2(Z\alpha)^5(m/M)m$ to the level shifts in muonium generated by the three gauge invariant sets of diagrams in Fig.~\ref{twoonepol}, Fig.~\ref{twoloppol} and Fig.~\ref{ee}

\beq \label{newresultmu}
\Delta E^{(Mu)}=-11.4308(2)\frac{\alpha^2(Z\alpha)^5}{\pi^3 n^3}\frac{m}{M}\left(\frac{m_r}{m}\right)^3m\delta_{l0}.
\eeq

Contribution to the Lamb shift of order $\alpha^7m$ in positronium  generated by the diagrams in Fig.~\ref{twoonepol} and Fig.~\ref{twoloppol} is given by the sum of the results in \eq{positrvac1} and \eq{positrvac2}

\beq \label{newresultpos}
\begin{split}
\Delta E^{(Ps)}&=
\Biggl(-\frac{4 \text{Li}_4\left(\frac{1}{2}\right)}{3}-\frac{19601 \zeta
   (3)}{10080}+\frac{476383}{544320}+\frac{289673 \pi ^2}{793800}\\
   & +\frac{7 \pi
   ^4}{2160}+\frac{1}{18} \pi ^2 \ln^22-\frac{\ln ^42}{18}-\frac{76}{315} \pi ^2 \ln2\Biggr)
\frac{\alpha^7m}{\pi^3n^3}\delta_{l0}\\
&=0.365117\ldots\frac{\alpha^7m}{\pi^3n^3}\delta_{l0}.
\end{split}
\eeq

\noindent
Numerically the contributions \eq{newresultmu} and \eq{newresultpos} are at the level of a few tenth of kilohertz and a few kilohertz, respectively. They are too small to be relevant for the results of the ongoing experiments. However, we expect that these corrections will become phenomenologically relevant in the future with further improvement of the experimental accuracy.

Calculation of corrections to the Lamb shifts in muonium and positronium generated by the diagrams in Fig.~\ref{twoonepol}, Fig.~\ref{twoloppol}  and Fig.~\ref{ee} is a step on the route to calculation of all spin-dependent and spin-independent hard corrections of order $\alpha^2(Z\alpha)^5(m/M)m$ in muonium and of order $\alpha^7m$ in positronium.  We hope to report results for the remaining hard contributions of this order in the near future.

\acknowledgments

Work of M.E. was supported by the NSF grants PHY-1724638 and PHY- 2011161.


\begin{thebibliography}{99}


\bibitem{mbb1982} F.~G.~Mariam, W.~Beer, P.~R.~Bolton et al., Phys. Rev.
Lett. \textbf{49}, 993 (1982).

\bibitem{lbd1999} W.~Liu, M.~G.~Boshier, S.~Dhawan et al., Phys. Rev. Lett. \textbf{82}, 711 (1999).

\bibitem{MuSEUM:2020mzm}
S.~Nishimura \textit{et al.} [MuSEUM],
Phys. Rev. A \textbf{104}, no.2, L020801 (2021)
[arXiv:2007.12386 [hep-ex]].


\bibitem{Kanda:2020mmc}
S.~Kanda, Y.~Fukao, Y.~Ikedo, K.~Ishida, M.~Iwasaki, D.~Kawall, N.~Kawamura, K.~M.~Kojima, N.~Kurosawa and Y.~Matsuda, \textit{et al.},
Phys. Lett. B \textbf{815}, 136154 (2021)
[arXiv:2004.05862 [hep-ex]].

\bibitem{apmghb1975} A.~P.~Mills, Jr. and G.~H.~Bearman, Phys. Rev. Lett. \textbf{34}, 246 (1975).

\bibitem{apmj1983} Allen~P.~Mills, Jr., Phys. Rev. A \textbf{27}, 262 (1983).

\bibitem{Ritter:1984mqy}
M.~W.~Ritter, P.~O.~Egan, V.~W.~Hughes and K.~A.~Woodle,
Phys. Rev. A \textbf{30}, no.3, 1331 (1984).

\bibitem{Sasaki:2010mg}
Y.~Sasaki, A.~Miyazaki, A.~Ishida, T.~Namba, S.~Asai, T.~Kobayashi, H.~Saito, K.~Tanaka and A.~Yamamoto,
Phys. Lett. B \textbf{697}, 121-126 (2011)
[arXiv:1002.4567 [physics.atom-ph]].

\bibitem{Ishida:2013waa}
A.~Ishida, T.~Namba, S.~Asai, T.~Kobayashi, H.~Saito, M.~Yoshida, K.~Tanaka and A.~Yamamoto,
Phys. Lett. B \textbf{734}, 338-344 (2014)
[arXiv:1310.6923 [hep-ex]].

\bibitem{Heiss:2018jbl}
M.~Heiss, G.~Wichmann, A.~Rubbia and P.~Crivelli,
J. Phys. Conf. Ser. \textbf{1138}, no.1, 012007 (2018)
[arXiv:1805.05886 [physics.atom-ph]].

\bibitem{Eides:2013yxa}
M.~I.~Eides and V.~A.~Shelyuto,
Phys. Rev. D \textbf{89}, no.1, 014034 (2014)
[arXiv:1311.1065 [hep-ph]].

\bibitem{Eides:2014nra}
M.~I.~Eides and V.~A.~Shelyuto,
Phys. Rev. Lett. \textbf{112}, no.17, 173004 (2014)
[arXiv:1402.5372 [hep-ph]].

\bibitem{Eides:2014xea}
M.~I.~Eides and V.~A.~Shelyuto,
Phys. Rev. D \textbf{90}, no.11, 113002 (2014)
[arXiv:1410.2930 [hep-ph]].


\bibitem{Eides:2000xc}
M.~I.~Eides, H.~Grotch and V.~A.~Shelyuto,
Phys. Rept. \textbf{342}, 63-261 (2001)
[arXiv:hep-ph/0002158 [hep-ph]].


\bibitem{Eides:2007exa}
M.~I.~Eides, H.~Grotch and V.~A.~Shelyuto,
Springer Tracts Mod. Phys. \textbf{222}, pp. 1-262 (2007)
doi:10.1007/3-540-45270-2.

\bibitem{Tiesinga:2021myr}
E.~Tiesinga, P.~J.~Mohr, D.~B.~Newell and B.~N.~Taylor,
Rev. Mod. Phys. \textbf{93}, no.2, 025010 (2021).

\bibitem{Eides:2018rph}
M.~I.~Eides,
Phys. Lett. B \textbf{795}, 113-116 (2019)
[arXiv:1812.10881 [hep-ph]].

\bibitem{Baker:2014sua}
M.~Baker, P.~Marquard, A.~Penin, J.~Piclum and M.~Steinhauser,
Phys. Rev. Lett. \textbf{112}, no.12, 120407 (2014)
[arXiv:1402.0876 [hep-ph]].

\bibitem{Adkins:2014dva}
G.~S.~Adkins and R.~N.~Fell,
Phys. Rev. A \textbf{89}, no.5, 052518 (2014)
[arXiv:1402.7040 [hep-ph]].

\bibitem{Eides:2014nga}
M.~I.~Eides and V.~A.~Shelyuto,
Phys. Rev. D \textbf{89}, no.11, 111301 (2014)
[arXiv:1403.7947 [hep-ph]].


\bibitem{Adkins:2014xoa}
G.~S.~Adkins, C.~Parsons, M.~D.~Salinger, R.~Wang and R.~N.~Fell,
Phys. Rev. A \textbf{90}, no.4, 042502 (2014)
[arXiv:1407.8232 [hep-ph]].

\bibitem{Eides:2015nla}
M.~I.~Eides and V.~A.~Shelyuto,
Phys. Rev. D \textbf{92}, no.1, 013010 (2015)
[arXiv:1506.00175 [hep-ph]].



\bibitem{Adkins:2015jia}
G.~S.~Adkins, M.~Kim, C.~Parsons and R.~N.~Fell,
Phys. Rev. Lett. \textbf{115}, no.23, 233401 (2015)
[arXiv:1507.07841 [hep-ph]].

\bibitem{Adkins:2015dna}
G.~S.~Adkins, C.~Parsons, M.~D.~Salinger and R.~Wang,
Phys. Lett. B \textbf{747}, 551-555 (2015)
[arXiv:1506.03835 [hep-ph]].

\bibitem{Adkins:2016btu}
G.~S.~Adkins, L.~M.~Tran and R.~Wang,
Phys. Rev. A \textbf{93}, no.5, 052511 (2016)
[arXiv:1603.03930 [hep-ph]].

\bibitem{Eides:2017uoy}
M.~I.~Eides and V.~A.~Shelyuto,
Phys. Rev. D \textbf{96}, no.1, 011301 (2017)
[arXiv:1705.09166 [hep-ph]].

\bibitem{Karshenboim:2003vs}
S.~G.~Karshenboim,
Int. J. Mod. Phys. A \textbf{19}, 3879-3896 (2004)
[arXiv:hep-ph/0310099 [hep-ph]].

\bibitem{Karshenboim:2005iy}
S.~G.~Karshenboim,
Phys. Rept. \textbf{422}, 1-63 (2005)
[arXiv:hep-ph/0509010 [hep-ph]].

\bibitem{Adkins:2018lvj}
G.~S.~Adkins,
J. Phys. Conf. Ser. \textbf{1138}, no.1, 012005 (2018).

\bibitem{Chu:1988} Steven~Chu, A.~P.~Mills, Jr., A.~G.~Yodh, K.~Nagamine, Y.~Miyake, and T.~Kuga,
Phys. Rev. Lett. \textbf{60}, 101 (1988).

\bibitem{Maas:1994} F.~E.~Maas, B.~Braun, H.`Geerds \textit{et al.}, Physics Letters A \textbf{187}, 247 (1994).

\bibitem{Meyer:1999cx}
V.~Meyer, S.~N.~Bagaev, P.~E.~G.~Baird, P.~Bakule, M.~G.~Boshier, A.~Breitruck, S.~L.~Cornish, S.~Dychkov, G.~H.~Eaton and A.~Grossmann, \textit{et al.},
Phys. Rev. Lett. \textbf{84}, 1136 (2000)
[arXiv:hep-ex/9907013 [hep-ex]].

\bibitem{Fan:2013qpk}
I.~Fan, C.~Y.~Chang, L.~B.~Wang, S.~L.~Cornish, J.~T.~Shy and Y.~W.~Liu,
Phys. Rev. A \textbf{89}, no.3, 032513 (2014)
[arXiv:1310.1660 [physics.atom-ph]].

\bibitem{Oram:1983sd}
C.~J.~Oram, J.~M.~Bailey, P.~W.~Schmor, C.~A.~Fry, R.~F.~Kiefl, J.~B.~Warren, G.~M.~Marshall and A.~Olin,
Phys. Rev. Lett. \textbf{52}, no.11, 910 (1984).

\bibitem{Woodle:1990ky}
K.~A.~Woodle, A.~Badertscher, V.~W.~Hughes, D.~C.~Lu, M.~W.~Ritter, M.~Gladisch, H.~Orth, G.~Zu Putlitz, M.~Eckhause and J.~Kane, \textit{et al.},
Phys. Rev. A \textbf{41}, 93-105 (1990).


\bibitem{Crivelli:2018vfe}
P.~Crivelli,
Hyperfine Interact. \textbf{239}, no.1, 49 (2018).
[arXiv:1811.00310 [physics.atom-ph]].

\bibitem{Ohayon:2021dec}
B.~Ohayon, Z.~Burkley and P.~Crivelli,
SciPost Phys. Proc. \textbf{5}, 029 (2021).


\bibitem{yksu2018} Y. Kuno and S. Uetake, talks at the
International Workshop on "Muonium and its Related
Topics" at Osaka University, December 11, 2018.


\bibitem{Ohayon:2021qof}
B.~Ohayon, G.~Janka, I.~Cortinovis, Z.~Burkley, L.~d.~Bourges, E.~Depero, A.~Golovizin, X.~Ni, Z.~Salman and A.~Suter, \textit{et al.},
[arXiv:2108.12891 [physics.atom-ph]].


\bibitem{Chu:1984zz}
S.~Chu, A.~P.~Mills and J.~L.~Hall,
Phys. Rev. Lett. \textbf{52}, 1689-1692 (1984).

\bibitem{Fee:1993zz}
M.~S.~Fee, A.~P.~Mills, S.~Chu, E.~D.~Shaw, K.~Danzmann, R.~J.~Chichester and D.~M.~Zuckerman,
Phys. Rev. Lett. \textbf{70}, 1397-1400 (1993).

\bibitem{Crivelli:2016fjw}
P.~Crivelli and G.~Wichmann,
[arXiv:1607.06398 [hep-ph]].

\bibitem{Mills:2016} A.~P.~Mills Jr., Adv. At. Mol. Opt. Phys. \textbf{65}, 265 (2016).

\bibitem{Cassidy:2018tgq}
D.~B.~Cassidy,
Eur. Phys. J. D \textbf{72}, no.3, 53 (2018).

\bibitem{Gurung:2020hms}
L.~Gurung, T.~J.~Babij, S.~D.~Hogan and D.~B.~Cassidy,
Phys. Rev. Lett. \textbf{125}, no.7, 073002 (2020).

\bibitem{Tomalak:2018uhr}
O.~Tomalak,
Eur. Phys. J. A \textbf{55}, no.5, 64 (2019)
[arXiv:1808.09204 [hep-ph]].

\bibitem{Eides:2009dp}
M.~I.~Eides and V.~A.~Shelyuto,
Phys. Rev. Lett. \textbf{103}, 133003 (2009)
[arXiv:0907.1923 [hep-ph]].

\bibitem{Eides:1992gr}
M.~I.~Eides, H.~Grotch and D.~A.~Owen,
Phys. Lett. B \textbf{294}, 115-119 (1992).

\bibitem{Eides:1994mm}
M.~I.~Eides and H.~Grotch,
Phys. Rev. A \textbf{52}, 1757 (1995)
[arXiv:hep-ph/9411220 [hep-ph]].

\bibitem{Kallen:1955} G.~Kallen and A.~Sabry, Kgl. Dan. Vidensk. Selsk. Mat.-Fis. Medd. 29 (1955) No.17.

\bibitem{Schwinger:1973} J.~Schwinger, Particles, Sources and Fields, Vol.2 (Addison-Wesley, Reading,
MA, 1973).

\bibitem{Eides:2000kj}
M.~I.~Eides, H.~Grotch and V.~A.~Shelyuto,
Phys. Rev. A \textbf{63}, 052509 (2001)
[arXiv:hep-ph/0012372 [hep-ph]].


\end{thebibliography}
\end{document}